\documentclass{elsart}
\usepackage{graphicx}

\newcommand{\be}{\begin{equation}}
\newcommand{\ee}{\end{equation}}
\newcommand{\bea}{\begin{eqnarray}}
\newcommand{\eea}{\end{eqnarray}}
\newcommand{\down}{\downarrow}
\newcommand{\up}{\uparrow}

\newcommand{\f}{\frac}

\begin{document}

\begin{frontmatter}
\title{Concurrence in the two dimensional XXZ- and transverse field Ising-models}
\author{Olav F. Sylju{\aa}sen}
\address{NORDITA, Blegdamsvej 17, DK-2100 Copenhagen {\O}, Denmark}
\ead{sylju@nordita.dk}

\begin{abstract}
Numerical results for the concurrence and
bounds on the localizable entanglement are obtained for 
the square lattice spin-1/2 XXZ-model and the transverse field Ising-model
at low temperatures using quantum Monte Carlo. 

\end{abstract}


\end{frontmatter}

While physicists have been dealing with entangled states
since the birth of quantum mechanics, one has only recently focused on
{\em quantifying} entanglement in general states of quantum systems.
 
Entanglement of two spin-1/2 spins in {\em pure} states is well-understood,
however there is no unique definition of entanglement in general {\em mixed} states.
One natural definition of entanglement of two spins in a mixed state is the entanglement of formation, $E_F$\cite{formation}. It is natural as it counts the minimum number of maximally entangled states needed to construct the state using a restricted set of rules. $E_F$ is a property of the reduced density matrix for the two spins in question and is a convex function, implying that it is not possible to create $E_F$ by mixing states classically. A less desirable property of $E_F$ is however that it doesn't capture the forms of entanglement as is present in for instance the GHZ state. Thus one can argue that $E_F$, although natural, is in some cases a too restrictive measure of entanglement.

An opposite extreme is the entanglement of assistance, $E_A$\cite{DiVincenzo}, which is quantified as the maximal two-spin entanglement achievable by performing  any types of measurements on the rest of the spins. $E_A$ is also a property
of the reduced density matrix, but is a concave function. It is a useful concept when entanglement concentration is needed, as for instance when building quantum repeaters. However to be of practical use the experimenter must be able to perform all kinds of measurement types in order to realize the potential of $E_A$.

To make this measure more practical the localizable entanglement, $E_L$, was recently introduced\cite{Verstraete}. It is similar to $E_A$ but only allowing
local measurements on the spins in the environment. $E_L$ is not a property of the reduced density matrix alone and as such can characterize more complicated features of the wave-function not captured by two-point correlation functions.
From a condensed matter viewpoint this form of entanglement may provide a 
useful concept. 
Traditionally states of many-body
systems are characterized by their two-point correlation functions, and
many important characteristics such as phases and phase transitions 
are obtained from these. However
there are important cases where exotic topological orders play important
roles requiring one to go beyond two-point correlation functions. Thus
it becomes vital to identify quantities that describe these orders. Recently
a connection between entanglement and topological order was established\cite{Verstraete2}. It was shown that the existence of the string order parameter in the AKLT-model, a topological quantity, is a direct consequence of the presence of long-range localizable entanglement.
While finding the exact magnitude of $E_L$ obviously require information beyond what is obtainable using two-point correlation functions, it is quite remarkable and interesting that it is possible to obtain {\em bounds} on its 
magnitude using purely two-point correlation functions\cite{Verstraete}. 

In
this Letter we will obtain numerical bounds on $E_L$ as well as calculate
the entanglement of formation for two general and 
important 2D quantum spin-1/2 models: The XXZ-model and the transverse field Ising-model. 

While a number of theoretical studies have been devoted to quantifying
entanglement in models of
{\em one dimensional} quantum magnets\cite{Gunlycke,Arnesen,Osterloh,Osborne,Wang}, materials behaving as two dimensional quantum magnets have in the last two decades been heavily studied because
of their connection to high-temperature superconductivity. 
Having
such well characterized materials makes them interesting also from an
applied quantum information theoretic point of view.
While it remains an open issue as to what extent quantum spin systems are actually useful for building quantum devices, a study of their entanglement properties constitutes a prerequisite.

For a pure state of two spins $|\psi \rangle = a|\down \down \rangle + b | \up \down \rangle + c | \down \up \rangle + d | \up \up \rangle$, the deviation from a product state can be simply expressed in terms of the concurrence $C = 2 |ad-bc|$.
The concurrence is functionally related to the pure state entanglement $E$ as defined by the Von Neumann entropy of one subsystem: $E=-x \log_2 x -(1-x) \log_2 (1-x)$, where $x = 1/2+\sqrt{1-C^2}/2$. 
In a remarkable paper Hill and Wootters\cite{HillWootters} showed that the same functional relation holds between $E_F$ and $C_F$, where $C_F$ is the concurrence generalized to a mixed state:
\be
C_F = {\rm max} \left\{ 0, \sqrt{\lambda_1}-\sqrt{\lambda_2} -\sqrt{\lambda_3} -\sqrt{\lambda_4} \right\}.
\ee
where $\lambda_1 \geq \lambda_2 \geq \lambda_3 \geq \lambda_4$ are the eigenvalues of $\rho \tilde{\rho}$. $\rho \equiv \rho_{ij}$ is the reduced density matrix of the spins $i$ and $j$ obtained by tracing over all other spins in the system
\be \label{densitymatrix}
 \rho_{ij} = \f{1}{4} \sum_{\alpha,\beta} 
                     \langle \sigma^\alpha_i \sigma^\beta_j \rangle
                     \sigma^\alpha_i \otimes \sigma^\beta_j,
\ee
the brackets denote ground state expectation values and $\sigma$ are the Pauli matrices. Greek (latin) indexes indicate spin (site) labels.
$\tilde{\rho}_{ij} = \sigma^y_i \otimes \sigma^y_j {\rho}^* \sigma^y_i \otimes \sigma^y_j$ is the time-reversed density matrix. 

There are no simple expressions for $E_A$. However an upper bound on $E_A$ is the concurrence of assistance $C_A$\cite{DiVincenzo} which 
is defined\cite{Laustsen} as
\be \label{concass}
C_A = {\rm max} \sum_k p_k C(\pi_k)
\ee
where $C$ is the pure state concurrence and $p_k$ is the probability of obtaining the pure two-spin state $\pi_k$ using a particular measurement procedure on all spins except spin $i$ and $j$. The maximum is taken over all possible measurement procedures. 
$C_A$ can be calculated as\cite{Laustsen}
\be \label{ca}
C_A = \sqrt{\lambda_1}+\sqrt{\lambda_2} +\sqrt{\lambda_3} +\sqrt{\lambda_4}.
\ee
$C_A$ ($C_F$) is the smallest (largest) concave (convex) function coinciding with the pure state concurrence for pure states\cite{Uhlmann}.

$C_A$ is also an upper bound on $E_L$ which is defined as in Eq.~(\ref{concass}( with the difference that the maximum is taken over {\em local} measurement procedures only. Note that because $E_L$ is defined in terms of the pure state concurrence the entropy of one subsystem $S(\rho_i={\rm Tr}_j \rho_{ij})$ which is an upper bound on $E_A$ is not necessarily an upper bound on $E_L$.
A lower bound on $E_L$ is gotten
by calculating the largest singular value of the $3 \times 3$ matrix of connected correlation functions\cite{Verstraete}:
\be \label{connected}
   Q^{\alpha \beta}_{ij} = \langle \sigma^\alpha_i \sigma^\beta_j \rangle
   -\langle \sigma^\alpha_i \rangle \langle \sigma^\beta_j \rangle.
\ee

The Hamiltonian for the XXZ-model is
\be
H_{\rm xxz} = \sum_{\langle i,j \rangle} \left\{
              -\left(  \sigma^x_i \sigma^x_j 
	                     + \sigma^y_i \sigma^y_j \right) + 
	      \Delta \sigma^z_i \sigma^z_j \right\},
\ee
where the sum is taken over all nearest neighbor sites on a square lattice
and units are chosen such that the exchange coupling is unity.
Using the symmetries of this Hamiltonian the concurrence can be written
\be \label{formation}
C_F =  \f{1}{2} \rm{max} \left\{ \f{}{} 0 ,  |\langle \sigma_i^x \sigma_j^x \rangle 
            + \langle \sigma_i^y \sigma_j^y \rangle | 
		   - 
		   \sqrt{ (1+ \langle \sigma^z_i \sigma^z_j \rangle )^2
		     - \langle \sigma^z_i + \sigma^z_j \rangle^2 }
		   \right\}.
\ee
This expression is strictly valid only when there is no spontaneous symmetry breaking. However as shown in Ref.~\cite{Olav} the concurrence is unaffected by 
spontaneous symmetry breaking for the zero-field XXZ-model, thus this expression yields the correct result.
The concurrence of assistance is
\be \label{assistance}
C_A =  \f{1}{2} \sqrt{ (1+ \langle \sigma^z_i \sigma^z_j \rangle )^2
		     - \langle \sigma^z_i + \sigma^z_j \rangle^2 }
      +
      \f{1}{2} \sqrt{ (1- \langle \sigma^z_i \sigma^z_j \rangle )^2
	- \langle \sigma^z_i - \sigma^z_j \rangle^2 }.
\ee

The (equal-time) correlation functions in Eqs.~(\ref{connected}),(\ref{formation}) and (\ref{assistance}) are calculated using Monte Carlo simulations
employing the Stochastic Series Expansion technique\cite{SSE} with directed-loop
updates\cite{SS} on a $L\times L$ square lattice with periodic boundary conditions. This technique utilizes a high temperature expansion to map the quantum system to a classical vertex model residing on a space-time lattice. This mapping has
many parallels with the Suzuki-Trotter mapping of a quantum $d$ dimensional system to a classical system in $d+1$ dimensions. The efficiency of the method relies on a fast motion thru configuration space which is achieved by the non-local directed loop updates leading to very short autocorrelation times. Correlation functions of observables diagonal in the 
representation basis can be read off directly from the configuration while correlation functions of off-diagonal observables can
be read off from how the loop propagates in space-time. This is because the
tail and the head of the loop act as operator insertions in the spin-configuration. This makes two-point correlations particularly easy to measure. It is possible to measure higher order correlators as well, but at considerably increased
computational cost. 

\begin{figure}
\begin{center}
\includegraphics[clip,width=10cm]{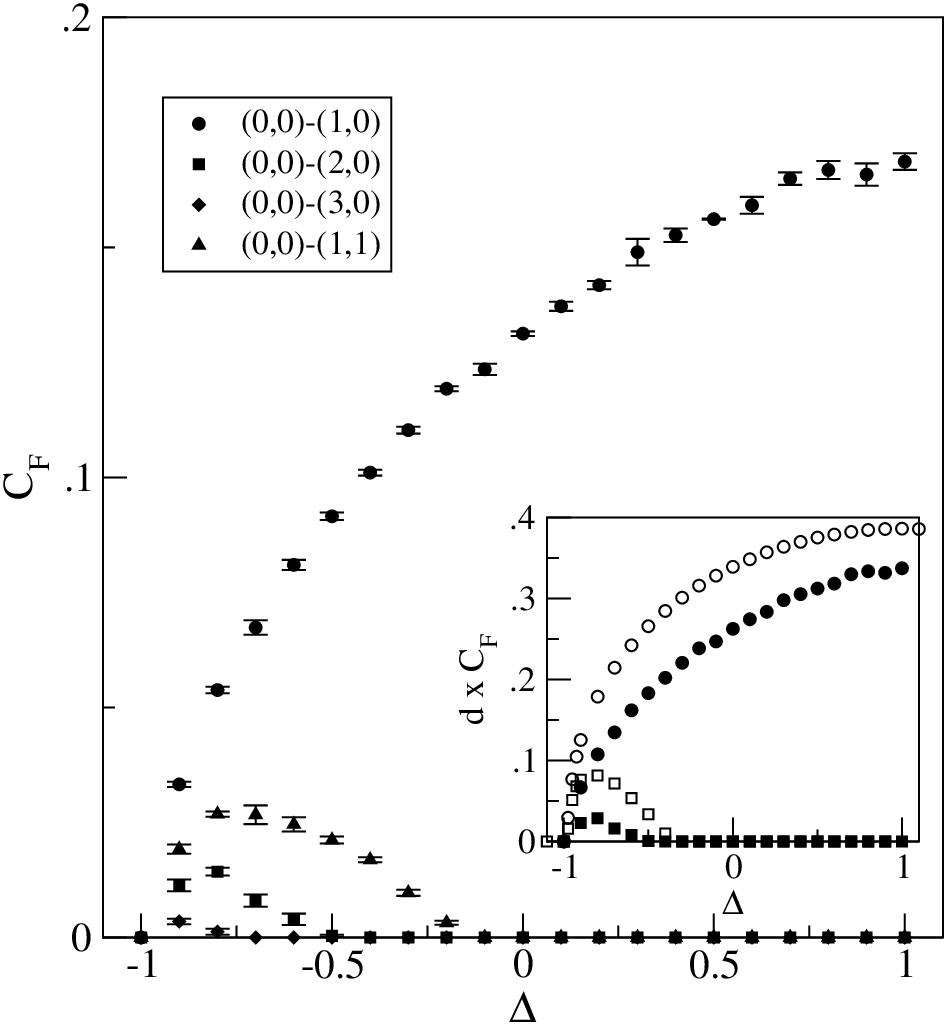}
\caption{The concurrence as function of $\Delta$ for different separations $(i_x,i_y)-(j_x,j_y)$ between sites $i$ and $j$ for a square lattice XXZ-model with $32 \times 32$ sites.
The inverse temperature is $\beta =80$. Inset is a comparison to the 1D result(open symbols, 128-site chain at $\beta=80$) for the linear nearest and next-nearest neighbor $C_F$ where the 2D $C_F$ is multiplied by d=2.} 
\label{xxz}
\end{center}
\end{figure}   

Fig.~\ref{xxz} shows $C_F$ plotted as a
 function of $\Delta$ for different separations between sites $i$ and $j$ at
a low temperature ($\beta =80$). Decreasing the temperature further did not change the results significantly. 
The nearest neighbor concurrence is the largest and peaks at the antiferromagnetic point where
it reaches a value $0.169 \pm 0.002$ which
is in good agreement with the value 0.1694 gotten from the ground state energy.
The concurrence for bigger distances between the spins reaches a maximum close to $\Delta=-1$ and decreases rapidly with increasing $\Delta$ and separation between $i$ and $j$.
The concurrence between diagonal neighbors is bigger than that between linear next-nearest neighbors. 
The general behavior resembles that of the concurrence in the XXZ-chain although the actual values here are lower. This is expected on grounds of 
the ``monogamy'' property of entanglement: The more neighbors a spin is entangled with, the less entanglement per pair. However even when correcting for the increased number of neighbors $C_F$ is lower for the square lattice than for the chain. This can be seen from the inset in Fig.~\ref{xxz}.

\begin{figure}
\begin{center}
\includegraphics[clip,width=10cm]{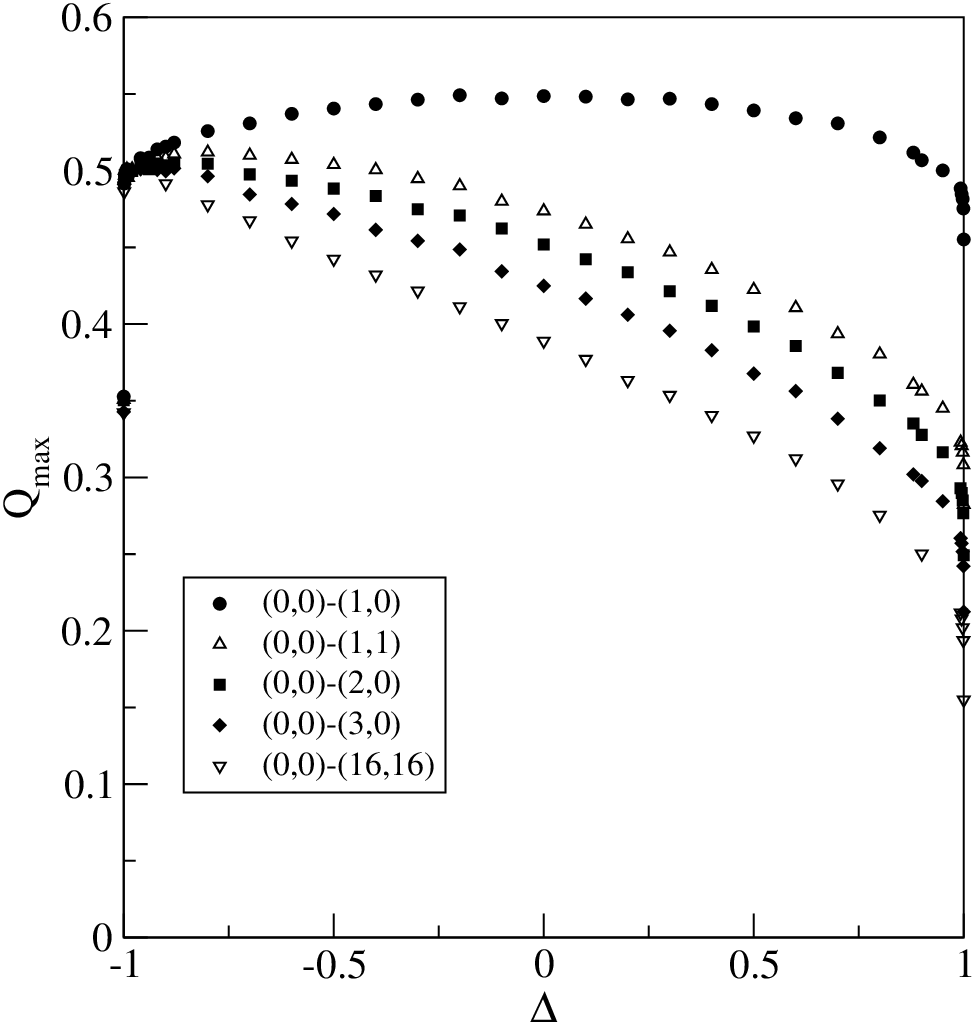}
\caption{Lower bound on $E_L$ as function of $\Delta$ for different separations between sites $i$ and $j$ for a square lattice XXZ-model with $32 \times 32$ sites. The inverse temperature is $\beta =80$.} 
\label{boundsXXZ}
\end{center}
\end{figure}

For $\Delta \le -1$, the ground state is a product state of spins aligned in the same direction, thus $C_A=0$, and all forms of entanglement are zero. 
For $|\Delta| \le 1$ the expectation value $\langle \sigma^z \rangle =0$ at any finite temperatures, thus $C_A=1$. A lower bound on $E_L$ is gotten from $Q^{xx}_{ij}$ which is the maximum correlation function in the regime $\Delta \le 1$. A plot of $Q^{xx}_{ij}$ as a function of $\Delta$ for different site separations is shown in Fig.~\ref{boundsXXZ}. 
For big separations $Q^{xx}_{ij}$ behaves as a power law at low temperatures and so decay slowly
 implying the existence of $E_L$ over long distances. 
The lower bound is largest towards the ferromagnetic point, where also
the decay with distance is very slow. Thus one expect to see the largest $E_L$ there. This is similar to the 1D situation where the lower bound decays as
$|i-j|^{-\arccos(-\Delta)/\pi}$ with a $\Delta$-dependent prefactor\cite{JinKorepin}, thus also decaying slowest at the ferromagnetic point.

 For $\Delta \geq 1$ the spin symmetry is broken by an infinitesimal staggered field and $C_A$ deviates from unity. It is
interesting to ask what $C_A$ is for infinite separation between $i$ and $j$ in the broken symmetry phase. Using $\langle \sigma^z_i \sigma^z_j \rangle = \langle \sigma^z_i \rangle \langle \sigma^z_j \rangle$ in the limit $|i-j| \to \infty$ one finds $C_A(|i-j| \to \infty) = 1-m_s^2$ , where $m_s = |\langle \sigma^z \rangle |$ is the staggered magnetization which at the Heisenberg point is $m_s(\Delta=1)=0.614$ and increases towards $m_s(\Delta \to \infty)=1$. 
It is interesting to note that true long-ranged $E_L$ is not ruled out by this upper bound at the Heisenberg point ($\Delta=1$) or at any finite $\Delta$, in contrast to the Ising case ($\Delta \to \infty$) where $C_A = 0$. 
The lower bound on $E_L$ decreases however rapidly to zero as the separation $|i-j|$ is increased. This is a consequence of the finite excitation gap in the spectrum for $\Delta \geq 1$.

The Hamiltonian for the transverse field Ising model is
\be
 H = - \sum_{\langle i,j \rangle} 
                          \lambda \sigma^x_i \sigma^x_{j} 
			  -\sum_i \sigma^z_i.
\ee
This model exhibits a phase transition at $\lambda=\lambda_c$. For
$\lambda \ll \lambda_c$ the transverse magnetic field dominates and the
ground state is similar to free spins in a magnetic field. For
$\lambda \gg \lambda_c$ , the system is like an Ising model without a magnetic
field thus having a doubly degenerate ground state. In an experimental situation this degeneracy is broken which must be taken into
account when calculating correlation functions. 

The concurrence can be expressed as
\bea
C_F & = & \f{1}{2} \mbox{max} \left( \f{}{} 0,
         | \langle \sigma^x_i \sigma^x_j \rangle
          +\langle \sigma^y_i \sigma^y_j \rangle |
	  -\sqrt{ (1+\langle \sigma^z_i \sigma^z_j \rangle)^2 - 
	          \langle \sigma^z_i + \sigma^z_j \rangle},
	  \right. \nonumber \\
 	 &   &  \left. | \langle \sigma^x_i \sigma^x_j \rangle
          -\langle \sigma^y_i \sigma^y_j \rangle |
	  -\sqrt{ (1-\langle \sigma^z_i \sigma^z_j \rangle)^2} \right)
\eea
This expression is also valid for $\lambda>\lambda_c$ regardless of symmetry breaking\cite{Olav}.

\begin{figure}
\begin{center}
\includegraphics[clip,width=8cm]{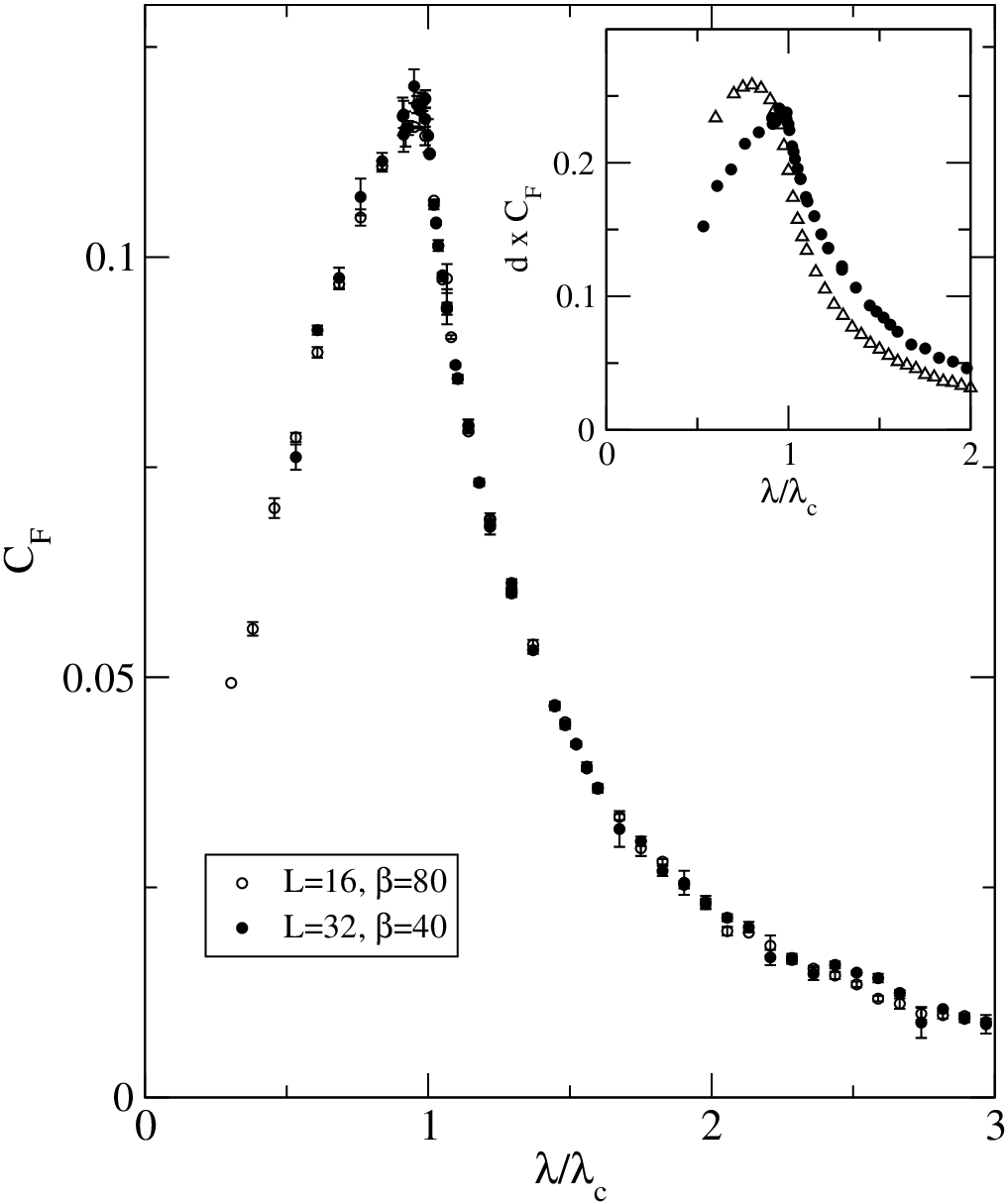}
\caption{Nearest neighbor concurrence for the transverse field Ising model on a square lattice.
The different curves are for different lattice sizes $L \times L$ at different temperatures. $\lambda_c=0.32841$ is gotten from Ref.~\cite{Hamer}. The inset
shows a comparison to the transverse Ising chain (open triangular symbols). In
the inset the 2D concurrence is multiplied by a factor $d=2$.}
\label{traIsing_plot}
\end{center}
\end{figure}   

The results for the nearest neighbor concurrence is plotted in Fig.~\ref{traIsing_plot}. The curve resembles largely that seen for the transverse 
field Ising-chain (seen in the inset) although the values are lower consistent with entanglement monogamy. However, in contrast to the one dimensional case where the concurrence peaks well on the high-field side of the transition, $\lambda_{\mbox{peak 1D}} \approx 0.8 \lambda_c$\cite{Osterloh}, the concurrence peaks here much closer to the critical point. The statistical errors makes it difficult to assess the exact location of the peak. 

\begin{figure}
\begin{center}
\includegraphics[clip,width=8cm]{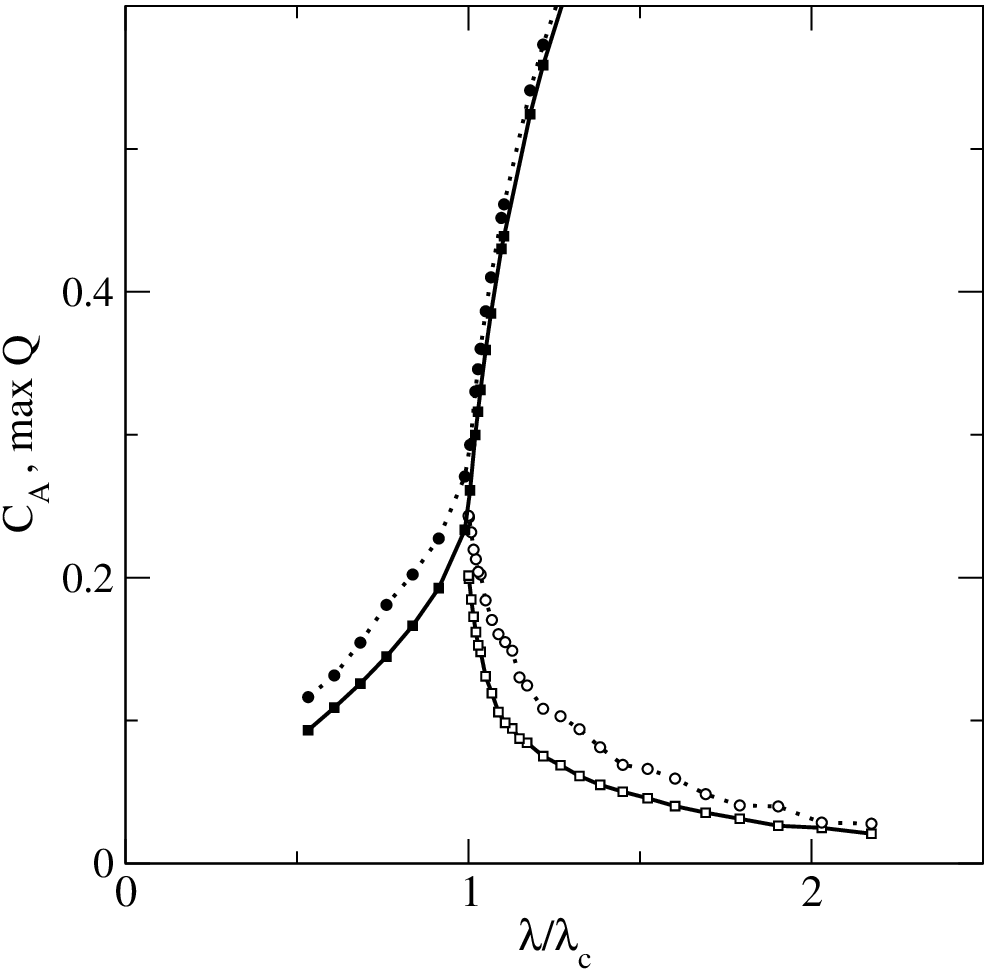}
\caption{Bounds on the nearest-neighbor $E_L$ on a $32 \times 32$ lattice for $\beta=40$. Upper bounds are indicated with circular symbols while square symbols indicate lower bounds. The lines are guides to the eye. 
The closed symbols are for a system without a symmetry breaking parameter, while the open symbols are obtained by adding a small longitudinal field ($H_x=0.001$).}
\label{Isbounds}
\end{center}
\end{figure}   

For $\lambda < \lambda_c$ Eq.~(\ref{assistance}) is valid. However
in the symmetry broken phase there are no simple explicit expression for $C_A$ so 
we use Eq.~(\ref{ca}) where the eigenvalues are gotten from numerically diagonalizing $\rho \tilde{\rho}$ taking care to measure also
$\langle \sigma^z \sigma^x \rangle$ correlators in $\rho$.

The bounds on $E_L$ is shown in Fig.~(\ref{Isbounds}). 
For $\lambda < \lambda_c$ the upper and lower bounds give relatively tight bounds on $E_L$. Both increase from zero and approach each other as $\lambda \to \lambda_c$. For $\lambda > \lambda_c$ the symmetric ground state approaches the GHZ state for $\lambda \to \infty$, thus $E_L$ $\to 1$. In this case the upper and lower bounds almost coincides thus giving very tight bounds on $E_L$. Results for the 
experimentally more realistic case when the symmetry is broken is also shown.
Here the upper and lower bounds approaches zero quite rapidly. For other distances between the spins, the lower bound drops rapidly to zero away from $\lambda_c$. The upper bound also decreases away from $\lambda_c$, but not as fast. These results resembles those obtained in 1D\cite{Verstraete}.

In conclusion the concurrence and bounds on the localizable entanglement was calculated numerically for the 2D XXZ-model and the transverse field Ising model. 
While the gross features of the results are similar to results in 1D,
it is interesting to note the slow decay of the localizable entanglement close to the ferromagnetic point in the XXZ-model as well as the
location of the maximum concurrence in the transverse field Ising model.

\ack{
Monte Carlo calculations were in part carried out using NorduGrid,
a Nordic facility for Wide Area Computing and Data Handling.
}

\end{document}